\documentclass[onecolumn,authoryear]{els-mrw} 

\usepackage{amsmath,amssymb,amsfonts,amsthm,makeidx,graphicx}
\usepackage{txfonts}
\usepackage{helvet}

\begin{document}

\chapter{Fast Radio Bursts}\label{chap1}

\author[1]{J. I. Katz}%


\address[1]{\orgname{Washington University}, \orgdiv{Department of Physics and McDonnell Center for the Space Sciences}, \orgaddress{St. Louis, Mo. 63130 USA}}


\maketitle


\begin{abstract}[Abstract]
	Eighteen years after their discovery, the astronomical sources and
	radiation mechanisms of fast radio bursts remain mysterious.
	Their radiation is as bright as that of pulsars, with brightness
	temperatures as high as $\sim 10^{36}$ K, implying coherent
	emission, but the plasma physics that forms the coherent charge
	bunches, with net charges of order a Coulomb, is not understood.
	Some FRB have been identified with galaxies at redshifts of a few
	tenths, but one originated within a globular cluster in the galaxy
	M81 at a distance of 3.6 Mpc.  A minority of FRB have been observed
	to repeat, in some cases thousands of times.  The vast majority of
	FRB have not been observed to repeat, but it is not known if they
	are truly ``one-offs'' or repeat at unobservably long intervals.
	Some FRB originate within dense, rapidly varying, plasma
	environments, while others appear to be surrounded by high vacuum.
	Hypotheses for their sources include magnetars and black hole
	accretion discs.
\end{abstract}


\seealso{Accretion Fundamentals, Magnetars, Black Hole Fundamentals}

\section*{Chapter Outline}
	\begin{itemize}
		\item Fast Radio Bursts---What are They?
		\item Discovery
		\item Properties
		\item Energetics
		\item Brightness
		\item Identifications
		\item Sources
		\item Repeaters {\it vs.\/} Non-Repeaters
		\item Emission Processes
		\item Environment
		\item Approaching the Uncertainty Principle
		\item Periodically Modulated Activity
	\end{itemize}

\begin{glossary}[Glossary]
\term{FRB} Fast Radio Bursts.

\term{Brightness Temperature} The temperature of a hypothetical black body
	that would produce an observed flux density.  For nonthermal sources
	this is not a physical temperature, and for coherent emitters like
	FRB and pulsars it may be enormous.

\term{Dispersion Measure (DM)} The column density of free electrons
	integrated along the line of sight to a radio source.  Lower
	frequency radiation is delayed compared to higher frequency
	radiation because its propagation speed is slowed by plasma
	refraction.

\term{Rotation Measure (RM)} The integral along the line of sight to a radio
	source of the product of the free electron density and the
	line-of-sight component of magnetic field.  If the source is
	linearly polarized its polarization angle varies with frequency at
	a rate proportional to RM.  RM is diagnostic of the near-source
	environment.

\term{Magnetars} Hypermagnetized neutron stars, with surface fields 
	in the range $10^{14}$--$10^{15}$ Gauss.  They are the sources of
	Soft Gamma Repeaters (SGR), and hypothesized to be the sources of
	FRB.

\term{Soft Gamma Repeater (SGR)} Neutron stars, believed to be magnetars,
	that have produced short (0.1--0.2 s) bursts of comparatively low
	energy (100 keV--few MeV) gamma rays.  An extremely energetic
	($10^{46}$-- $10^{47}$ ergs) burst was followed by large numbers of
	very much less energetic bursts; hence ``Repeater''.

\term{Curvature Radiation} The radiation emitted by relativistic charged
	particles, such as electrons, positrons or clumps of particles with
	unneutralized net charge, accelerated by their motion along a curved
	magnetic field line.

\term{Coherent Radiation} The radiation emitted by an accelerated
	unneutralized clump of charged particles, with power proportional to
	the square of the net charge in the clump.

\end{glossary}

\section*{Objectives}
\begin{itemize}
	\item Fast Radio Bursts were a surprising discovery
	\item Fast Radio Bursts are extremely bright
	\item Some Fast Radio Bursts repeat, others are not known to repeat
	\item The mechanism and sources of Fast Radio Bursts are unknown
\end{itemize}

\section{Introduction}\label{introduction}
Fast Radio Bursts (FRB) are distant enigmatic astronomical events whose
sources and physical mechanisms remain uncertain (and controversial) 17 years
after their discovery in 2007 \cite{L07}.  FRB have been observed at
frequencies from 110 MHz to 8 GHz, but extensive searches have found no
counterparts at frequencies from visible light to gamma-rays.

FRB resemble radio pulsars in some respects, and were first discovered in
archival pulsar search data.  Like pulsar pulses, FRB are brief, with
durations typically a few milliseconds but as short as $\sim 10\,\mu$s.
Like pulsars, FRB have extraordinarily high brightness temperatures $T_b =
{\cal O}(10^{36}\,$K).  Again like pulsars, FRB have varying and diverse
polarization, ranging from nearly 100\% linearly polarized to nearly 100\%
circularly polarized.

However, unlike pulsars, FRB do not have known counterparts at frequencies
outside the radio bands.  FRB differ from pulsars in other respects---some
FRB have been observed to repeat, but none have shown the stable periodic
repetition that characterizes pulsars and identifies them as rotating
magnetized neutron stars.

A fraction of repeating FRB have been localized to external galaxies (it is
difficult to obtain accurate coordinates for a source not observed to
repeat), typically at redshifts $z$ of a few tenths, although one was found
in the neighboring galaxy M81 at a distance 3.6 Mpc.  A few of these FRB
localized to galaxies are associated with persistent (steady) radio sources,
although it is not known if these persistent sources also radiate the
bursts or are unrelated properties of the host galaxies.  In contrast, all
known pulsars are Galactic, mostly at distances of a few kpc or less.  We
may say that most FRB are at ``cosmological'' distances, and that some are
suggested (on the basis of the column density of intervening intergalactic
plasma) to be at redshifts $z = {\cal O}(1)$.  But the directly measured
redshifts imply comparatively nearby cosmological distances, in contrast to
quasars, whose density in the universe peaked around $z=2$ and have been
observed up to $z=7.64$.

There are several reasons FRB are fascinating objects.  First, is simply
that they were unexpected, with extraordinary properties, and remain
unexplained.  Second, there is no agreement as to the astronomical nature of
their sources---finding them in galaxies tells us nothing about their
sources because galaxies contain every known (and likely some unknown)
classes of astronomical object.  Third, although they are clearly not
pulsars, their brightness temperatures are in the same class as those of the
brightest pulsars.  In both cases, this implies coherent emission---charges
clump (in FRB, to net charges of order a Coulomb!) and radiate together,
with a radiated intensity proportional to the square of the number of
elementary charges in the clump.  The plasma physics of how Nature does this
remains obscure, 57 years after the discovery of pulsars, but rather than
requiring special circumstances, coherent emission is widespread in 
astronomy, occurring in planetary magnetospheres, long period radio
transients (a subset of polars, synchronously rotating magnetic white dwarf
binaries polars) and stellar coron\ae.  FRB are an extreme example.

For a recent review article on FRB, with emphasis on theory and models, see
\cite{Z23}.  Technological advances in electronics have enabled base-band
observations (measuring radiation electric fields and electromagnetic wave
phases, rather than only the radiation spectrum) at GHz frequencies, which
promise to reveal the plasma processes emitting the observed radiation; see
\cite{CHIME24} for a recent review of FRB observations.
\section{Discovery}\label{discovery}
\begin{figure}
	\centering
	\includegraphics[width=\columnwidth]{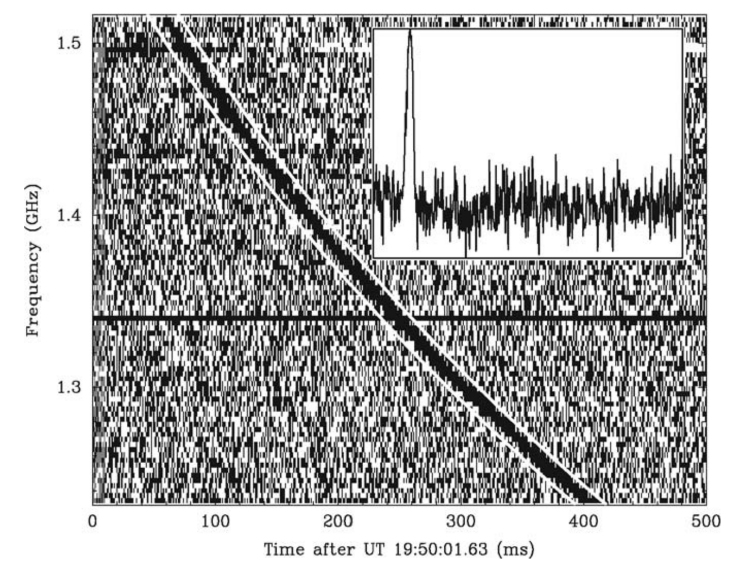}
	\caption{\label{L}Discovery of the first Fast Radio Burst:  the
	received intensity of the ``Lorimer burst'' as a function of radio
	frequency and time.  This observation was made at the Parkes
	Observatory in Australia, a 64 m diameter radio telescope.  The data
	are only one bit deep---if the spectral flux density was above a
	threshold the narrow rectangle corresponding to that time and
	frequency is black, if below threshold it is white.  The threshold
	was chosen so that detector and sky noise is above threshold in 
	about half the time-frequency elements and below threshold in the
	other half.  The dark curved path shows the burst.  Its curvature
	indicates propagation delay in plasma between the emitter and the
	observer; higher frequencies have higher group velocity and arrive
	earlier.  The adjoining white curves were added to emphasize the
	signal.  The horizontal black line shows a narrow spectral band
	excluded because of electronics failure.  The inset shows the
	frequency-integrated flux as a function of time, where the
	frequency-dependent (dispersion) delays have been removed; the
	intrinsic full width at half maximum of the burst was about 7 ms.
	Reproduced with permission from \cite{L07}.}
\end{figure}

The first FRB was discovered \cite{L07} in 2007 in archival data in a search
for pulsars.  The radiation from the FRB was actually received in 2001, but
the data were only analyzed six years later; how many other discoveries are
waiting to be uncovered in archives?  The discovery data are shown in
Fig.~\ref{L}.

Radio astronomy is plagued with ``radio-frequency interference'', emissions
from anthropogenic terrestrial sources (in Fig.~\ref{L} a horizontal black
like blanks out detected interference), so the first response to this
discovery was skepticism.  The astronomical community did not generally
accept that FRB are actual astronomical events until their discovery was
confirmed by the discovery of several more FRB in 2012.  In fact, emission
resembling FRB was detected a few years later that turned out to be
leakage from microwave ovens when their doors were opened.  Although these
``perytons'' were quickly discredited (and how a microwave oven excited by a
2.45 GHz cavity magnetron could produce 1--1.5 GHz radiation was never
explained), they were a reminder of the problem of anthropogenic
interference in radio astronomy.

The difference between the arrival times of higher and lower frequency
radiation from the same source is described by its Dispersion Measure (DM)
which is proportional to the integrated column density of electrons along
the path of radiation:
\begin{equation}
	\label{DM}
	\tau(\nu) - \tau(\nu = \infty) = {e^2 \over 2 \pi m_e c \nu^2}
	\int\!d\ell\,n_e,
\end{equation}
where $d\ell$ is the length element along the path, $e$ is the electron
charge, $m_e$ the electron mass, $c$ the speed of light and $\nu$ the
frequency of the radiation.  The integral in Eq.~\ref{DM} is the Dispersion
Measure, usually (for convenience) given in units of parsec-cm$^{-3}$ rather
than cm$^{-2}$.  This had long been familiar to pulsar astronomers, who
immediately recognized plasma dispersion as the cause of the later arrival
of lower frequency radiation in Fig.~\ref{L}.

Fitting to the data in Fig.~\ref{L} showed $\text{DM} = 375\,$pc-cm$^{-3}$.
This would be extraordinary for a pulsar; most pulsars have DM of tens of
pc-cm$^{-3}$, with the few exceptions having long path lengths in the
ionized gas of the Galactic plane, and especially near the Galactic Center.
Yet the Lorimer burst had a Galactic latitude of $-42^\circ$, far out of the
Galactic plane.  This was clearly an extraordinary and novel source.  Its
DM was naturally attributed to intergalactic plasma, implying a distance of
hundreds of Mpc and a redshift of a few tenths.  Once its reality was
confirmed by the subsequent discovery of other FRB, a new branch of
astronomy was born.
\section{Properties: Summary}\label{properties}
About 800 distinct FRB have been observed as of August 2024.  They are very
heterogeneous (although perhaps not as heterogeneous as gamma-ray bursts,
about which it is said ``If you have seen one gamma-ray burst, you have seen
one gamma-ray burst.''), but a summary is possible:
\begin{itemize}
\item FRB are extremely bright, with brightness temperatures (defined in
	Sec.~\ref{brightness}) as large as $10^{36}\,$K.  If they weren't
	this bright, they would be undetectable at their great distances.
\item As of August, 2024, 56 FRB have been observed to repeat
	\begin{itemize}
		\item More than 1000 repetitions have been observed from a
			few of these repeaters
		\item Much more is known about these ``repeaters'' because
			it is possible to point a telescope at their known
			locations and collect extensive data
		\item Apparent non-repeaters may repeat at unobserved long
			intervals, but the distribution of repetition rates
			is bimodal.
	\end{itemize}
\item FRB durations range from tens of $\mu$s to tens of ms.
\item FRB Dispersion measures range up to about 2000 pc-cm$^{-3}$
	\begin{itemize}
		\item It is not known how much of the dispersion measures
			are attributable to intergalactic plasma (this
			component is directly related to redshift according
			to cosmological theory) and how much to denser
			plasma near the FRB source.
		\item Inferring distance from DM is therefore uncertain
	\end{itemize}
\item Many FRB have complex sub-pulse structure.
\item No FRB has been observed to emit periodic pulses (unlike pulsars,
	which are defined by their periodic pulses), but two FRB have shown
	that they emit bursts only within periodically recurring several-day
	stretches of activity.
\item The spectra of FRB are much narrower than the spectrum of radiation
	emitted by an accelerated point charge.  This indicates the
	radiating charge distribution has spatial structure, nearly periodic
	at the radiated wavelength (allowing for Doppler shifts).
\item FRB polarizations vary, some nearly 100\% linearly polarized and others
	nearly 100\% circularly polarized.
\item Some FRB have linear polarization with position angle varying with
	frequency, indicative of Faraday rotation in magnetized plasma along
	the line of sight, quantified by the rotation measure (RM).
\item In most repeating FRB the observed DM and RM are not observed to vary,
	consistent with a high vacuum environment.
\item In a few repeating FRB variations in DM and and RM are observed, from
	which a near-source plasma density ${\cal O}(10^8)\,$cm$^{-3}$ and
	magnetic field as large as $\sim 17\,$mGauss have been inferred,
	implying a dense chaotic plasma environment.
\item At least one FRB has shown scintillation, indicating a source region
	of dimension $\lesssim 10^{10}\,$cm.
\end{itemize}
\section{Energetics}\label{energetics}
How energetic are Fast Radio Bursters?  This is critical in distinguishing
among models and theories, for the first question to be asked is ``Is this
model capable of producing the observed energy?''.

The energy requirements of a FRB source are not easy to quantify.  A strong
FRB might have a fluence (time-integral of flux) of 10 Jy-ms, where the
Jansky (Jy) is defined as $10^{-26}\,$W/(m$^2$-s-Hz).  Integrating over a
nominal 1 GHz bandwidth, this is $10^{-19}\,$J/m$^2$.  For a source at 3 Gpc
distance (redshift $z \sim 1$) emitting isotropically, the total energy
radiated would be about $10^{40}$ ergs, with an instantaneous power of about
$10^{43}$ ergs/s.  

These are formidable numbers; the instantaneous power would be about
$10^{-3}$ of the luminosity of a bright quasar, or more than $10^9 L_\odot$.
However, they are almost certainly misleading, because FRB almost certainly
do not radiate isotropically.  Very likely, they are narrowly beamed, and
we only observe those that are (fortuitously) beamed in our direction.

Unfortunately, we do not know how narrowly FRB are beamed.  If radiated by
electrons with Lorentz factors $\Gamma$, they instantaneously illuminate
about $1/\Gamma^2$ steradians; for a plausible $\Gamma \sim 100$ (promoting
curvature radiation in a neutron star's inner magnetosphere to GHz
frequencies) the solid angle illuminated would be about $10^{-4}$ steradian,
multiplying the energy and instantaneous power by a factor of $10^{-5}$.
Of course, a wide bundle of relativistic emitters would broaden the
illuminated solid angle.

Beaming reduces the energy and power requirements of individual bursts, but
does not reduce the mean radiated power unless we are (implausibly) in a
preferred direction with respect to the source.  If bursts are emitted in
random directions isotropically distributed about the source, the mean
radiated power is the nominal (``isotropic-equivalent'') power of an
individual burst multiplied by the burst duty cycle (Sec.~\ref{duty}).
For active repeaters this might be as much as $10^{38}\,$erg/s, and several
orders of magnitude less for apparent non-repeaters.  The mean power would
further reduced if active repeaters preferentially emit bursts in our
general direction, rather than isotropically.

Beaming does not reduce the mean power emitted by all the FRB sources in the
universe, or the mean energy density of FRB radiation in the universe.  This
constrains theoretical models, but so far has not been constrained by
observation.
\section{Brightness}\label{brightness}
The most striking characteristic of FRB (and also of pulsars) is their
brightness temperature $T_b$.  This is defined as the temperature a black
body would have it if emitted the same flux $F_\nu$ as the observed FRB,
where we take the Wien ($h\nu \ll k_B T$) limit, where $k_B$ is the Boltzmann
constant, of the Planck function:
\begin{equation}
	\label{Tb}
	k_B T = {F_\nu c^2 \over 2 \nu^2}.
\end{equation}
$F_\nu$ has the units of erg/(s-cm$^2$-Hz-sterad), where the solid angle of
emission is dimensionless, but is essential.

In order to evaluate $T_b$ it is necessary to estimate the solid angle
subtended at the observer of the emitting region.  This is cannot be
measured directly, but is usually estimated as $(c \Delta t/D)^2$, where
$\Delta t$ is the width of the emitted burst or pulse, $c \Delta t$ is an
estimate of the transverse size of the emitter and $D$ is the distance to
the observer (for a relativistically expanding cloud of emitters the
transverse size may exceed $c \Delta t$ by a factor $\Gamma^2$, where
$\Gamma$ is the Lorentz factor of expansion).  These estimates are very
uncertain, but, both for pulsars and FRB, indicate $T_b$ as large as
${\cal O}(10^{36})\,$K.

$T_b$ is not a physical temperature.  Eq.~\ref{Tb} indicates spectacularly
high $T_b$ for an ordinary radio transmitter, even though no part of it is
hot, much less moving relativistically.  Large $T_b$ indicate coherent
emission, in which the electromagnetic radiation emitted by $N$ elementary
charges adds coherently; in effect, the radiating charge is $N e$ rather
than $e$, and the power radiated is $\propto N^2$.  In some FRB $N e$ may be
${\cal O}(1\,$ Coulomb), implying $N = {\cal O}(10^{19})$.  Coherent
emission requires that, if the charges move nonrelativistically, they be
confined to a region of size $\ll \lambda$, where $\lambda$ is the
wavelength of the emitted radiation, and that if they move relativistically
they be confined to a region of length $\ll \lambda\Gamma^2$ in their
direction of motion.  In a 1 MHz radio antenna $N$ might be
${\cal O}(10^{13})$ and the net charge ${\cal O}(1\,\mu$Coulomb).

Narrow-band emission may be produced if the radiating charge distribution
is not narrowly clumped (in which case, aside from amplitude, it radiates
the same extremely broad spectrum as a single point charge \cite{Jackson}),
but is periodic in space.  For example, if a series of relativistically
moving charge clumps move along the same curved path (as in synchrotron or
curvature radiation) an observer in the direction of their motion will see a
radiation field that peaks every time a charge clump moves in its direction
as it follows the curved path.  If there are a large number of equally
spaced clumps, the received signal will be periodic with peaks spaced by the
intervals between the appearance of the clumps moving towards the observer
(when the emission of relativistically moving charges peaks).  The existence
of such narrow band spectra tells us something about the spatial
distribution of the radiating charge density, and hence about the plasma
physics that creates that periodic charge distribution.
\section{Identifications}
As of August, 2024, 56 FRB have been identified with host galaxies.
Unsurprisingly, these are mostly FRB with comparatively low dispersion
measures (typically a few hundred pc-cm$^{-3}$).  This indicates
comparatively small distances (at which a host galaxy would be brighter,
facilitating identification), with redshifts of a few tenths, although a
nearby FRB would have a large dispersion measure if it were embedded in a
dense cloud of ionized gas.

The importance of these identifications is that it is straightforward to
measure the redshift, and hence distance, of a host galaxy by optical
spectroscopy.  With a sufficient number of identifications, as now exist,
it is possible to estimate the distribution of FRB sources in the universe.
This may be compared to the distribution of hypothetical source objects.
For example, statistical studies have compared the distribution of FRB in
the universe to its star formation history.  Some, but not all, such studies
have indicated that the density of FRB peaks several billion years after the
peak of star formation.  This suggests that FRB are not produced by young or
short-lived (massive) stars or other phenomena associated with young stars
(such as most types of supernov\ae\ and young neutron stars).  This is a
difficulty for models in which FRB are produced by short-lived products of
massive stars, such as young magnetars (hypermagnetized neutron stars, found
in Galactic supernova remnants), whose active lifetimes appear to be
$\lesssim 10^4\,$y.  Subsequent slow evolution, whether of individual
objects or of populations, may be necessary to produce the sources of FRB.

Another inference from identification of FRB with galaxies whose redshifts
can be measured is that at least a few FRB are embedded in gas clouds with
dispersion measures of several hundred pc-cm$^{-3}$.  This demonstrates that
a measured DM only sets an upper bound on the FRB's distance because an
unknown fraction, but possibly the majority, of the measured DM could be
produced by passage through plasma local to the FRB and not through
intergalactic plasma.  By itself, this doesn't tell us much about the
immediate environment of the FRB because ``local'' could be anything from
${\cal O}(10^{13}\,$cm) (or even less) to the dimensions of the host galaxy
or an associated halo ${\cal O}(10^{23}\,$cm).  Further information may be
inferred for a few FRB in which the dispersion measure is observed to vary;
see Sec.~\ref{environment}.
\section{Sources}\label{sources}
The most basic question to ask about FRB is what astronomical objects make
them.  17 years after their discovery, and 11 years after the general
acceptance of their reality, this question is still unanswered.  This is
unlike the history of pulsars (with which FRB have significant
similarities), whose nature as rotating magnetized neutron stars was
recognized, and generally appreciated within months of their discovery,
or of quasars (active galactic nuclei) whose nature as accreting
supermassive black holes was appreciated within a year.  It is closer to
the history of gamma-ray bursts, recognized in 1973 (in data originally
recorded in 1967) and understood in outline about two decades later.

The reason it was so easy to identify pulsars as magnetic neutron stars was
their emission of pulses at very steady, but slowly slowing, periods.
Their periodicity results from their rotation, as their beams sweep across
the sky like a lighthouse beacon (although possibly as fan, rather than
cone, beams).  As radiation drains energy from the rotation it gradually
slows, and from the rate of slowing their magnetic field may be inferred.
This picture was clear within months of their discovery.
\subsection{Magnetars}\label{magnetars}
Because of the similarities (brightness, coherent emission at radio
frequencies) between pulsars and FRB it was natural to look for similar
sources: neutron stars.  FRB are orders of magnitude more energetic, so
theorists turned to ``magnetars'', hypermagnetized neutron stars with fields
in the range $10^{14}$--$10^{15}$ Gauss, in comparison to the
$10^8$--$10^{13}$ Gauss fields of pulsars.  Magnetars were hypothesized in
1982 to explain Soft Gamma Repeaters (although at the time these were still
called gamma-ray bursts, and their spin-down in their quiescent phase called
Anomalous X-ray Pulsars hadn't yet been observed because of insufficient
instrumental sensitivity).  Magnetars can emit ample energy---the most
energetic Soft Gamma Repeater emitted more than $10^{47}$ ergs, several
orders of magnitude greater than even the highest estimates of FRB energies.
The fact that no detailed model of how a magnetar could make a FRB existed
was not a significant objection because, even a half-century after their
discovery, there was no such detailed model for the coherent emission of
pulsars.

Magnetars are known to rotate with periods in the range 2--12 seconds; this
is observed in the X-ray emission of their quiescent state.  Their spin
gradually slows, from which their magnetic fields are calculated, confirming
the extraordinary fields required to energize their Soft Gamma Repeater
outbursts.

Unfortunately for the magnetar explanation of FRB, some of its predictions
have not been borne out by observation.  Most importantly, extensive studies
of repeating FRB, involving thousands of bursts, have never found any
evidence that the bursts are periodic (the activity of two FRB sources is
periodically modulated with periods of many days, but the bursts themselves
are not periodic; see Sec.~\ref{modulation}).  It is hard to see how the
radiation---or any influence---of a rotating magnetized object could not
vary periodically with its rotational period.  Pulsars, and accreting
neutron star X-ray sources, are examples of this variation.  It is not
subtle---pulsars produce narrow pulses that accurately repeat at their
rotational periods, and the emission of known magnetars, in their quiescent
``Anomalous X-ray Pulsar'' phases, varies periodically at their rotational
periods with fractional amplitudes of several tens of percent or more.

Another issue for the magnetar model is that known magnetars occur at the
center of young ($\lesssim 10^4\,$y) supernova remnants that are evacuated
(by expansion of the supernova debris) to very low density (not directly
measured, but estimated theoretically to have an electron density $n_e \ll
1\,$cm$^{-3}$).  There is evidence (Sec.~\ref{environment}) that at least
one FRB is embedded in plasma many orders of magnitude denser than that.
This argument is not iron-clad, because FRB might be produced by very young
magnetars, whose surrounding supernova remnant hasn't expanded to the radius
of those enclosing much older Galactic magnetars, but as known repeating FRB
age without ceasing to emit bursts (one has been observed to emit bursts,
without evident diminution of activity, over a dozen years, setting a lower
bound on its lifetime) this loophole gradually becomes less plausible.
\subsection{Where?}\label{where}
If FRB are energized by magnetars, where is the radiation we observe
emitted?  Even for pulsars, that question has not been definitively
answered; we know that their energy is drawn from the neutron star's
rotational energy, but not whether the radiation is emitted in the inner
magnetosphere (for example, within a stellar radius of the neutron star's
surface), near the ``radius of light cylinder'' where a particle co-rotating
with the magnetic field would nominally be moving at the speed of light,
or elsewhere?

For magnetar models of FRB, possible sites of emission range from the
inner magnetosphere out to the surrounding supernova remnant.  For example,
it has been proposed that a magnetar burst (perhaps resembling a Soft
Gamma Repeater outburst, whose gamma-rays would be undetectably faint at
the $\sim$Gpc distances of FRB) send a jet of relativistic particles into
the supernova remnant.  This jet's interaction with the remnant's magnetic
field might radiate a FRB by a process known as synchrotron maser emission,
or perhaps by some other plasma process (plasma physics in strong fields is
notoriously difficult and complicated, and theory is unlikely to provide a
definitive answer).  This hypothesis would not resolve the problem that the
bursts of repeaters are not periodic; even if the radiation were produced
far from the magnetar, the interaction of a particle beam with surrounding
matter would be periodic at the magnetar's rotational period, like the
illumination of a distant ship or land by a lighthouse beacon.

A very recent study \cite{N24} interpreted the spectrum of a comparatively
close FRB (redshift $z=0.015$) as the result of scintillation, the
scattering of radiation in turbulent plasmas, analogous to the twinkling of
visible starlight.  As is familiar from stellar twinkling and the fact that
planets do not twinkle, strong fluctuations, and in this case sensitive
dependence of the intensity on frequency, imply a small angular size of the
source.  The conclusion was that the source region was no greater than about
$6 \times 10^9\,$cm across (about five times the diameter of the Earth) at a
distance of about 60 Mpc or $2 \times 10^{26}\,$cm, corresponding to an
angular width of $3 \times 10^{-17}\,$radian, or 30 attoradians (0.03
femtoradian)!  If confirmed for other FRB, this excludes models involving
interactions of beams with matter distant from the source of the beam, such
as the hypothesized synchrontron masers driven by explosive outbursts of
magnetars.
\subsection{What Else?}\label{whatelse}
If FRB are not produced by magnetars, then by what else?  From their short
durations and the scintillation observations described in Sec.~\ref{where}
we know that their source regions must be small, although the uncertain
Lorentz factors $\Gamma$ of emitting particles may make this difficult to
quantify.  The energy of FRB is also difficult to estimate because their
emission is likely highly collimated (a general property of radiation by
relativistic particles), but to be observed at cosmological distances, even
at redshifts of a few tenths, they must be fairly energetic.  That points to
compact objects that, because of their deep gravitational potential wells
and (in the case of magnetars) high magnetic fields, may provide high energy
densities and emitted power.

An important constraint is the fact that some FRB repeat.  At least for
those FRB, catastrophic events (supernov\ae, neutron star mergers, neutron
star formation, tidal stellar disruption events$\ldots$) that destroy their
progenitors may be excluded.  In fact, there are not enough known
catastrophic events to explain even the apparently non-repeating FRB
\cite{H20}.

Another hypothesis is that FRB are produced in the throats of accretion
discs around black holes, perhaps black holes of intermediate mass
($10^2$--$10^6\,M_\odot$) \cite{K17}.  Such black holes, with masses between
those of binary black holes in our Galaxy and those observed by LIGO/VIRGO
as gravitational wave sources when they merge (both $\lesssim 30 M_\odot$)
and the supermassive black holes at the centers of active galactic nuclei,
have long been the subject of speculation, but have never been convincingly
observed.

Accretion discs around black holes are believed to encompass low-density
throats analogous to the whirlpool around a bathtub drain.  The low plasma
density in such a vortex (centrifugal acceleration forces rotating matter
away from the rotational axis, giving it a small scale height in the
direction perpendicular to the vortex wall) might permit the plasma
processes necessary to accelerate energetic particles and concentrate them
into the coherent bunches necessary to emit the coherent radiation of FRB.
Observations related to this hypothesis are discussed in
Sec.~\ref{modulation}.  
\section{Repeaters {\it vs.\/} Non-Repeaters}\label{rvsnr}
Some FRB repeat; others have never been observed to repeat.  The latter are
called ``apparent non-repeaters'' because we do not have definitive evidence
that they have never repeated in the past or that they will never repeat in
the future.  Are these two categories of FRB qualitatively different, or
only quantitatively different, with differing repetition rates?

This question has not been definitely answered.  There is evidence that
these two categories of FRB differ qualitatively.  That does not prove that
apparent non-repeaters have never and will never repeat, but only that there
are few intermediate cases.
\subsection{Duty Cycles}\label{duty}
One such difference occurs in their ``duty cycles'' $D$, the fraction of
time they emit radiation strong enough to be detected (this term originates
from radar electronics, where transmitters are either ``on'' or ``off'', and
$D$ is the fraction of time they are ``on'').  This can be generalized to
sources whose intensity $I$ varies as $D = \langle I \rangle^2/\langle I^2
\rangle$, where the brackets indicate time averages.  If certain plausible
statistical assumptions are satisfied $D$ is independent of the sensitivity
of the observing radio telescope.  The duty cycles of very active repeating
FRB are typically ${\cal O}(10^{-5})$ (a 1 ms burst roughly once per 100 s,
on average).  Upper limits of $10^{-10}$--$10^{-8}$ (no bursts in
$10^5$--$10^7$ s following a detected 1 ms burst) have been set on the duty
cycles of the best-observed apparently non-repeating FRB.  Of course, these
numbers depend on instrumental sensitivity, but the comparison between 
active repeaters and apparent non-repeaters is valid if observations of the
same sensitivity are compared.  This doesn't demonstrate that there is not a
continuum of less active repeaters with intermediate duty cycles, but does
indicate qualitative differences.
\subsection{Dynamic Spectra}\label{dynamicspectra}
\begin{figure}
                \centering
                \includegraphics[width=\columnwidth]{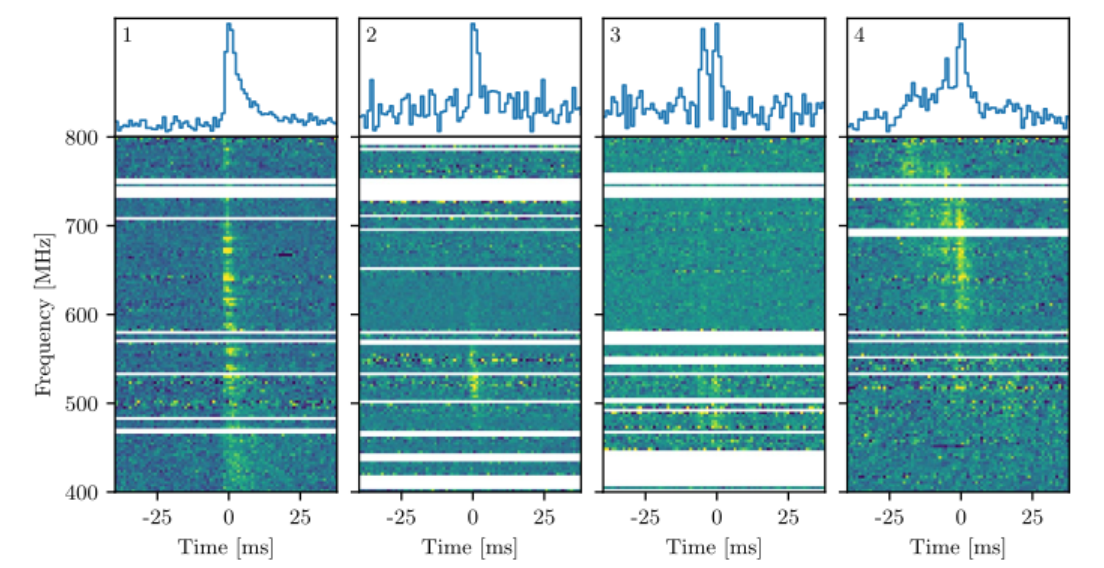}
                \caption{\label{P3} De-dispersed dynamic spectra of
                bursts of four types, reproduced with permission from
		\cite{P21b}.  The first three sub-figures are apparent
		non-repeaters.  The fourth is a repeater, and shows a drift
		to lower frequency, distinct from dispersion.  This ``sad
		trombone'' effect is found for most repeating bursts, {but
		is unusual for apparent non-repeaters}.}
        \end{figure}

\begin{figure}
                \centering
                \includegraphics[width=\columnwidth]{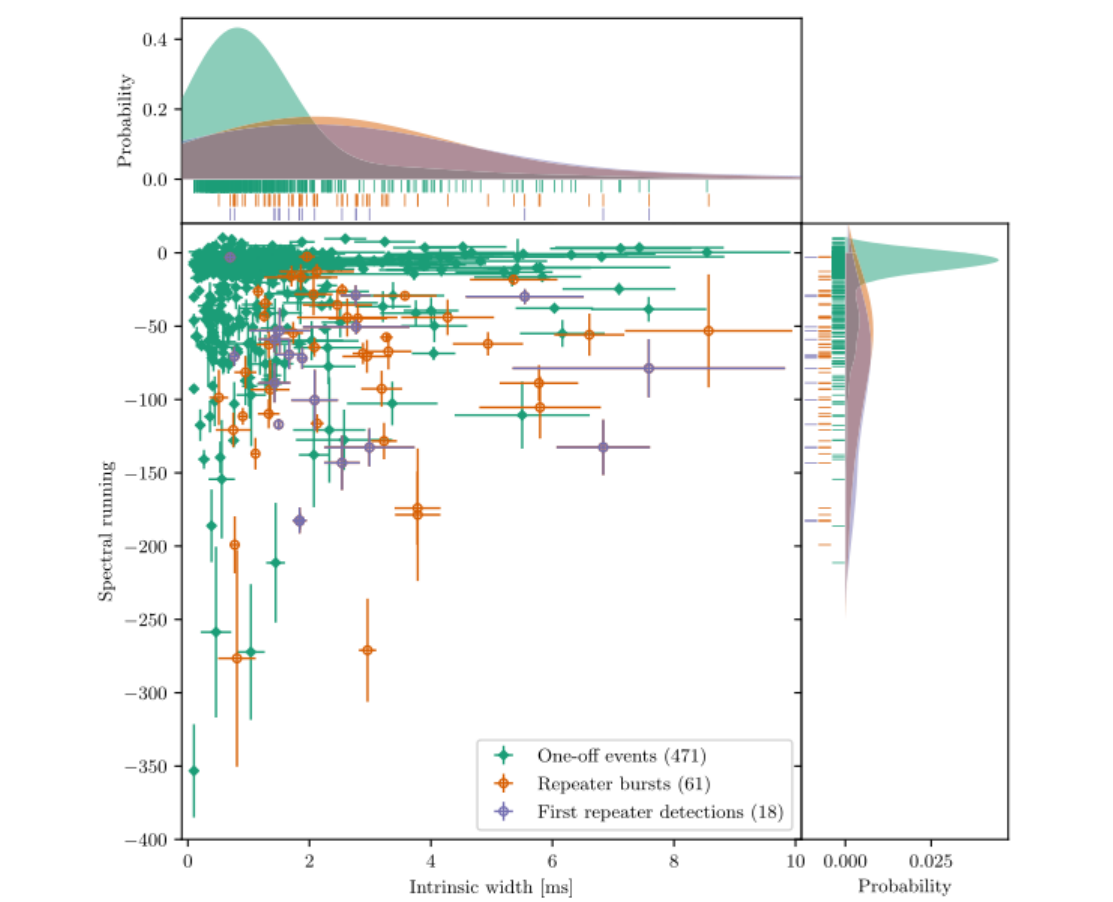}
                \caption{\label{P4} \bf Fitted spectral parameter ``running''
                {\it vs.\/} burst length for several hundred bursts (numbers
                 in figure), with apparent non-repeaters (``one-offs'')
		 green, repeaters purple and orange.  The distributions of
		 each variable, separately, overlap, but in the
		 two-dimensional space there is a clear separation.
		 Reproduced with permission from \cite{P21b}.}
        \end{figure}

\begin{figure}
                \centering
                \includegraphics[width=\columnwidth]{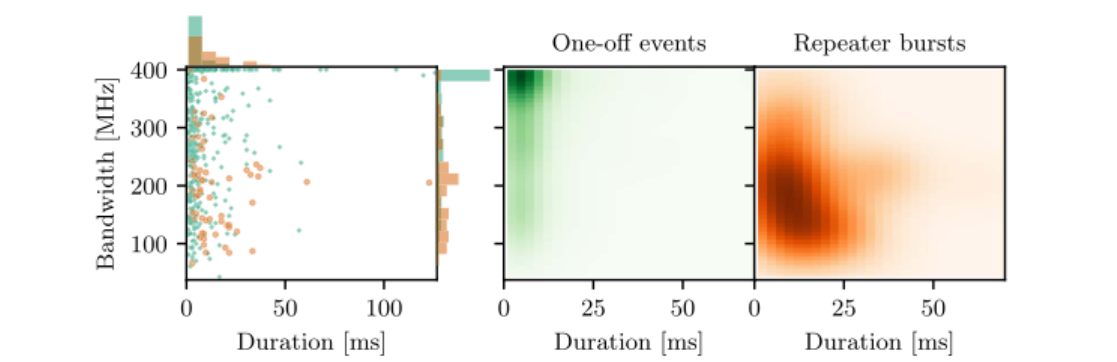}
                \caption{\label{P5} \bf Burst bandwidth {\it vs.\/}
		duration, with apparent non-repeaters (``one-offs'') green,
		repeaters purple and orange.  The distributions in the
		two-dimensional space are nearly disjoint.  Reproduced with
		permission from \cite{P21b}.}
        \end{figure}

Examination of individual bursts shows other qualitative differences.  For
example, Figs.~\ref{P3}--\ref{P5} compare the radio spectra of repeating and
apparent non-repeating FRB after dispersive time delays have been removed.
\section{Emission Processes}\label{emission}
What physical process emits the radiation we observe as a Fast Radio Burst?
Electromagnetic radiation is emitted when charges are accelerated.  This
can happen when they pass near other charges and are deflected
electrostatically (bremsstrahlung radiation), when radiation of one
frequency is scattered by an electron (Compton scattering), from internal
motions within atoms or nuclei, when electrons gyrate around a magnetic
field (cyclotron radiation if the electrons are moving non-relativistically
or synchrotron radiation if they are relativistic) or when electrons are
guided by curved magnetic field lines (curvature radiation).  Positrons emit
in the same manner, and with the same intensity (all else being equal) as
electrons, but radiation by protons or nuclei is generally negligible
because their large masses limit their acceleration and Lorentz factors
(with the exception of radiating nuclear or collisional gamma rays).

Of these processes, only synchrotron radiation in very small magnetic fields
and curvature radiation in larger magnetic fields can produce significant
emission at radio frequencies.  Plausible sources of the high power, short
pulse radiation of FRB must be compact, because a short time scale implies a
compact source region and the combination of compactness and high power
requires a high energy density in the source region.  The high brightness
(Sec.~\ref{brightness}) of FRB requires that the radiating charges are
strongly bunched.  This has suggested two classes of FRB emission processes,
neither of which has been developed into a quantitative model---the
necessary plasma physics is too hard.
\subsection{Synchrotron Maser}
We know that magnetars (Sec.~\ref{magnetars}) can emit enormous energy in
short bursts.  These bursts aren't as short as FRB---the peak emission of
SGR lasts 0.1--0.2 s, in comparison to the tens of $\mu$s to several ms of
FRB, but peak SGR emission rises in $< 1 ms$ (only upper limits of several
hundred $\mu$s exist, rather than actual measurements of this rise),
offering hope that shorter time scales may be involved.

How might a magnetar outburst create a FRB?  There is good evidence that a
SGR itself does not produce a FRB because a fortuitous unrelated radio
observation during the December 27, 2004 outburst of SGR 1806$-$20 set
strict upper bounds on any simultaneous FRB \cite{TKP16}.  However, a SGR
outburst is expected to expel relativistic plasma.  A young SGR (SGR
1806$-$20 is several thousand years old and its environs were swept clear by
its expanding supernova debris) would be surrounded by a young supernova
remnant.  The interaction of the relativistic plasma with the remnant plasma
has been hypothesized to create a ``synchrotron maser"---radio frequency
synchrotron radiation growing exponentially as a result of interaction
between two interpenetrating plasma streams.  In order that the radiation
occur at observed FRB frequencies 0.1--10 GHz, the magnetic field must
be small, the actual value depending on the Lorentz factor of the
relativistic debris but consistent with an expanded supernova remnant.
Although this hypothesis has difficulty explaining the inference
(Sec.~\ref{where}) that the emitting region in a FRB is very much smaller
($3 \times 10^10\,$cm) than even a year-old supernova remnant, the
exponential growth of the maser instability offers a possible means of
creating the coherent plasma structures required to emit coherent, high
brightness, radiation.
\subsection{Curvature Radiation}
Another means of producing radiation at the 0.1--10 GHz frequencies of
observed FRB might be the curvature radiation emitted by charge bunches
moving along the curved magnetic field lines of a neutron star or
(Sec.~\ref{whatelse}) black hole accretion funnel.  The observed angular
frequency of the radiation peaks around $0.3 \Gamma^3 c/r$, where $r$ is the
radius of curvature of the field line and $\Gamma$ the Lorentz factor of the
charge bunch (or corresponding to the phase velocity of a plasma wave).  If
$r \sim 10\,$km (describing the inner magnetosphere of a neutron star), then
1 GHz radiation is obtained for $\Gamma \sim 200$, a plausible value, or
roughly twenty times higher for an accretion vortex around an intermediate
mass black hole.

The sensitivity of the frequency to $\Gamma$ as well as the intrinsic
breadth of the spectrum radiated by a point change enables this hypothesis
to explain the wide range of frequencies seen in FRB from a comparatively
small range of $\Gamma$.  The narrow frequency structure of individual FRB
is then attributed to the convolution of the narrow-band spatial periodic
structure of the charge density excited by a plasma wave in any one FRB (or
at any particular time in that FRB).  This is consistent with the ability to
emit over a broad range of frequencies.
\section{Environment}\label{environment}
The environment of FRB is related to the nature of the source object.  For
example, a young magnetar is surrounded by a young supernova remnant that
may add a significant contribution to its dispersion measure, a contribution
that would systematically decline as the supernova remnant expands.  An
older supernova remnant would contribute negligibly.  An accreting object
might be in a chaotic environment with a more complicated time dependence.

In addition to dispersion measure, it is also possible to measure the
rotation measure of a linearly polarized FRB: the polarization angle of a
linearly polarized electromagnetic wave rotates as it propagates through a
magnetized plasma, and the cumulative rotation angle is proportional to the
line integral of the product of the line-of-sight component of magnetic
field and the electron density times the square of the wavelength.  The
wavelength dependence means that the rotation measure may be inferred from
the dependence of the linear polarization angle on frequency (assuming the
radiated polarization angle is independent of frequency, as it is for
synchrotron and curvature radiation).  If the FRB repeats, variations in
dispersion and rotation measures from burst to burst describe changes in
the environment of the FRB source.

Rotation measures are quadratic in quantities (magnetic field and plasma
density) that are expected to be tiny in the intergalactic medium.  Unlike
dispersion measures, that are likely mostly intergalactic, rotation measures
are heavily weighted to the immediate environment of a FRB source, and are
therefore sensitive to changes in that neighborhood.  Some FRB have rotation
measures hundreds of times greater than those of almost any Galactic
pulsars, indicating extraordinarily dense and strongly magnetized (by
astronomical standards) environments.

It might be assumed that the environment of a remote astronomical object
varies only slowly, perhaps undetectably slowly, but this does not turn out
to be the case.  In fact, FRB environments vary dramatically and rapidly.
Some repeating FRB have rotation measures that have changed sign \cite{AT23}
(remaining more than hundred times greater than those of Galactic pulsars)
in less than a year.  If the dispersion measure remains constant a changing
rotation measure may be explained by geometric realignment of the magnetic
field, but in a few FRB both rotation and dispersion measures have been
observed to change over the same period of time.  The ratio of the change is
rotation measure to that of dispersion measure then indicates the magnitude
of the magnetic field $B$ in the changing region (because rotation measure
is proportional to $B n_e \ell$ and dispersion measure is proportional to
$n_e \ell$, where $\ell$ is a characteristic length).

In FRB 20121102A the inferred magnetic field is estimated to be $\approx
17\,$mGauss \cite{K21a}, an extraordinary value.  Typical Galactic
interstellar magnetic fields are estimated to be ${\cal O}(3\,\mu$Gauss),
thousands of times less, so that the magnetic energy density near the FRB is
$\sim 10^7$ times greater than typical Galactic magnetic energy densities.
This FRB, and likely others, are embedded in extraordinarily dense and
chaotic magnetoionic environments.

\begin{figure}
                \centering
                \includegraphics[width=\columnwidth]{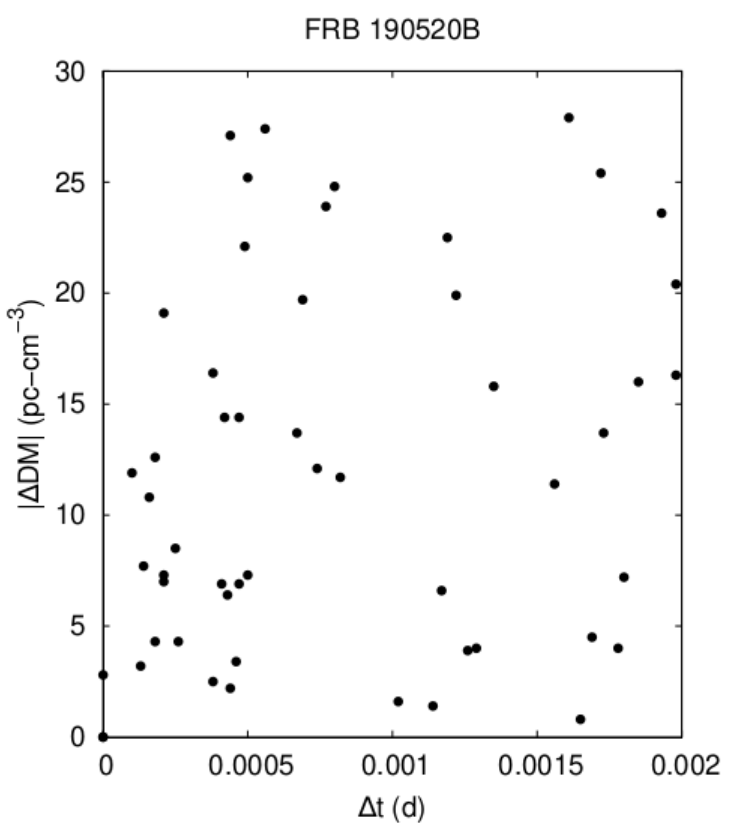}
                \caption{\label{190520B} $|\Delta\text{DM}|$ {\it vs.\/}
                $\Delta t$ for intervals between successive bursts of FRB
                20190520B on MJD 59373; data from Table S3 of \cite{AT23}.}
        \end{figure}

Even more remarkable are the variations of dispersion measure of FRB
20190520B, shown in Fig.~\ref{190520B}.  Its dispersion measure varied by
as much as 30 pc-cm$^{-3}$ on time scales of tens of seconds.  Interpreting
this as the result of clouds of plasma moving into and out of the line of
sight implies electron densities of ${\cal O}(10^9\,$cm$^{-3})$ even if the
plasma clouds are moving almost relativistically.  This suggests a dense
accreting environment (Sec.~\ref{whatelse}) rather than a magnetar in an
expanding supernova remnant.
\section{Approaching the Uncertainty Principle}\label{modes}
The recent development of base-band receivers recording the radiation
electric field on time scales short compared to the radiation frequency has
led to surprising results that pose challenges for the plasma physics that
creates the radiating charge bunches.  Observations of FRB 20220912A
\cite{H23} revealed microshots with temporal widths $\Delta t \le 31.25\,$ns
in a spectral channel of width $\Delta \nu = 16\,$MHz ($\Delta \omega =2\pi
\Delta \nu \approx 10^8\,$s$^{-1}$).  The product $\Delta \omega \Delta t
\approx 3$.  The uncertainty principle, which is a mathematical consequence
of the definition of the frequency spectrum as the Fourier transform of the
wave amplitude as a function of time, is written $\Delta \omega \Delta t \ge
1$ (with a small degree of freedom in the quantitative definitions of
$\Delta \omega$ and $\Delta t$); these observations and an earlier
observation of a Crab pulsar nanoshot approach the uncertainty principle
limit.

These observations are telling us something about the radiation processes
in these microshots and nanoshots.  The value of $\Delta \omega \Delta t$ is
the number of orthogonal electromagnetic wave modes that contribute
substantially to the radiation field; Nature has somehow arranged this
almost as well as an engineered single-mode laser.  Each orthogonal
electromagnetic wave mode is produced by an orthogonal plasma wave mode.  If
an instability makes the plasma wave modes grow exponentially, the
observation that only a few modes contribute most of the radiated power
requires that they outgrew almost all of the enormous number of modes
possible in a region of space.  This sets a lower bound on the number of
exponentiations of unstable growth from the origin of the plasma waves in a
thermal plasma \cite{K24}.
\section{Periodically Modulated Activity}\label{modulation}
One of the most important constraints on possible FRB models is that the
bursts are not periodic.  In contrast to pulsars, whose intervals between
pulses are the same to high accuracy (very slowly increasing as the neutron
star's rotation slows over thousands or millions of years), the intervals
between FRB pulses are randomly distributed.  These distributions are well
fit by log-normal functions, but that may only be a consequence of the fact
that log-normals are very flexible, capable of fitting a wide range of
empirical distributions.  However, the width of a fitted log-normal function
is a useful dimensionless parametrization of a distribution.

\begin{figure}
        \centering
        \includegraphics[width=\columnwidth]{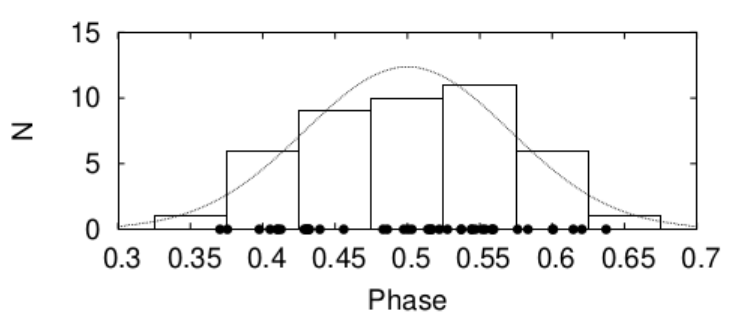}
        \caption{\label{jitter} Phases of the 44 individual bursts of FRB
        20180916B observed by \cite{Mck23a} for the fitted period of 16.315
        d (large dots), a bar graph of their distribution (solid) and the
	fitted Gaussian (dotted).  If the phase jitter results from a random
	walk about exact periodicity, the observed Gaussian distribution is
        expected.  The mean phase is defined as 0.5 and its standard
	deviation is 0.071.  Data from \cite{Mck23a}.}
\end{figure}

However, at least two FRB have shown a different kind of periodicity.
All observed pulses from FRB 20180916B occur within five day windows that
repeat with a 16.35 day period, and a similar modulation is observed with
a 160 day period for FRB 20121102A.  The most recent collection of data for
FRB 20180916B is shown in Fig.~\ref{jitter}.

Several different explanations of this periodicity have been suggested,
including the free rigid-body precession of a magnetar source, a very slowly
rotating neutron star, a binary orbit in which a companion or its wind
shadows or envelops a neutron star source of FRB, and emission of FRB along
the rotational axis of an accretion disc that precesses as a result of the
gravitational torque of a companion star that feeds the disc.  None of these
explanations has been demonstrated or compellingly disproven, nor (except
for the last) are the same mechanisms observed in other astronomical
objects.

The precessing disc hypothesis is based known precessing discs, of which
those in Her X-1 and SS 433 are the best known, with precession periods
(35 d and 164 d, respectively) comparable to those observed in the two
periodically modulated FRB.  Modulation of observed FRB activity results if 
FRB are emitted in directions near, but distributed with random offsets of a
few degrees around, the disc axis; when the axis points approximately in our
direction we may observe FRB, but when it precesses away from our direction
FRB activity is unobservable to us.  Similar random offsets (jitter) are
observed of the jets in SS 433.
\section{Conclusion}
Fast Radio Bursts were a discovery as unexpected and perhaps as
revolutionary as the discovery of pulsars.  After more than a decade of FRB
observation, it remains uncertain what astronomical objects make them, in
what environments they are found and their radiation mechanisms are
mysterious.  FRB are heterogeneous: some repeat, some apparently don't
repeat, some are found in dense, turbulent, strongly magnetized environments
while some appear to be in clean vacuum environments.  They are unified in
two ways: their brightness temperatures are extreme (like those of pulsars)
but their bursts are not periodic (unlike pulsars').  We hope that more and
improved observations from telescopes like CHIME/FRB, FAST and the Square
Kilometer Array will resolve these mysteries.
\bibliographystyle{Harvard}
\bibliography{reference} 

\end{document}